\documentclass[runningheads,a4paper]{llncs}
\usepackage[dvipdfmx]{graphicx}
\usepackage{amssymb}
\usepackage{amsmath}
\newcommand{\definedas}{\ \Leftrightarrow\ }
\newcommand{\always}{\Box}
\newcommand{\eventually}{\Diamond}
\newcommand{\equals}{\mathop{\texttt{=}}}
\title{%
Declarative Semantics of the Hybrid Constraint Language HydLa
\thanks{\footnotesize\addtolength{\baselineskip}{0pt}%
This is an English, author version of the paper that originally
appeared in the journal \textit{Computer Software}, Vol.~28, No.~1
(2011), pp.~306--311, 
doi:10.11309/jssst.28.1\_306,
\hfill\break
\texttt{https://www.jstage.jst.go.jp/article/jssst/28/1/28\char`_1\char`_1\char`_306/\char`_article/}.
When citing this paper, please cite the original, journal version with
a pointer to this English translation.\hfill\break
\textit{Notice for the use of this material}: The copyright of this
material is retained by the Japan Society for Software Science and
Technology (JSSST) as well as the authors, but JSSST is not
responsible for any inaccuracies of the translation.
This material is published on the web with the agreement of the
JSSST. Please comply with Copyright Law of Japan if any users
wish to reproduce, make derivative work, distribute or make available
to the public any part or whole thereof.}}

\titlerunning{Declarative Semantics of HydLa}
\author{Kazunori Ueda\inst{1} \and
        Hiroshi Hosobe\inst{2} \and Daisuke Ishii\inst{3}}
\institute{Department of Computer Science and Engineering, Waseda University
 \and National Institute of Informatics
 \and University of Nantes}

\authorrunning{Kazunori Ueda et al.}
\begin{document}
\maketitle

\begin{abstract}
Hybrid systems are dynamical systems with continuous evolution of
states and discrete evolution of states and governing equations.
We have worked on the design and implementation of HydLa, a
constraint-based modeling language for hybrid systems, with a view to
the proper handling of uncertainties and the integration of simulation
and verification.  HydLa's constraint hierarchies facilitate the
description of constraints with adequate strength, but its semantical
foundations are not obvious due to the interaction of various language
constructs.  This paper gives the declarative semantics of HydLa and 
discusses its properties and consequences by means of examples.
\end{abstract}

\addtolength{\baselineskip}{0pt}%
\section{Introduction}\label{sect:introduction}

Hybrid systems are dynamical systems with continuous evolution of
states and discrete evolution of states and governing equations.  We
have been developing a modeling framework of hybrid systems based on
the notion of Constraint Programming.  Our goal is to establish a
constraint-based paradigm in which (i) to describe diverse phenomena found
in physical, cyber-physical, and biological systems using logical
formulae involving equations and inequations and (ii) to solve or verify 
them using search techniques represented by constraint propagation.

Our motivation has been to establish, in the field of hybrid systems,
a declarative programming paradigm that directly handles
as source programs high-level description of
problems in mathematical and logical formulas, as opposed to
traditional formalisms based on automata and Petri Nets
\cite{HA}\cite{HPN}.  A similar approach was first taken by Hybrid CC
\cite{GJSB95}, and we have made a lot of experiments on Hybrid CC
programming.  However, we found that it was not necessarily 
straightforward to
specify constraints that a system consisting of alternate discrete and
continuous phases should satisfy, and this lead us to design a new
language that enables a concise description of hybrid systems.

Since the basic design of HydLa was established in 2008 \cite{Ueda},
we studied the details of the language through the description of a
number of examples \cite{Hirose}, developed a simulation
algorithm \cite{Shibuya} and a prototype implementation, and
explored technologies for implementing discrete changes
with guaranteed accuracy \cite{Ishii}.  All those studies
contributed to the clarification of the essence and subtle points of the
HydLa language specification.
Based on those experiences, this paper formulates
the declarative semantics of the core of HydLa, and discusses its
descriptive power and properties by means of examples.

\section{Overview of HydLa}\label{sect:overview}

HydLa is a declarative language for hybrid systems.  Its objective
is to allow one to provide the mathematical formulation of a given
problem with minimal modification and to simulate or analyze them.
For the design principles and related work of HydLa, the readers
are referred to \cite{Ueda}.

Dynamical systems that HydLa aims to handle are in general
represented as a countable number of real-valued functions $x_1(t)$,
$x_2(t)$, \dots $(t\ge 0)$ that include integer-valued functions
as a special case.
A HydLa program imposes constraints on the behavior of 
those functions (hereafter called \textit{trajectories}) that may cause 
continuous or discrete changes over time.  The declarative semantics
of a HydLa program $P$ is defined as a satisfaction relation between
trajectories $\overline{x}(t)=\{x_i(t)\}_{i\ge 1}$ and $P$, or
equivalently, the set of all $\overline{x}(t)$'s that satisfy $P$.

In order to describe hybrid systems in a concise manner,
the use of hierarchies to 
represent \textit{defaults} and \textit{exceptions}
will
play an important role exactly as in knowledge representation and 
object-oriented design.  Consider a ball bouncing on a floor.  The
change of the velocity of the ball is determined by the gravity
most of the time (default), while it is determined by the collision
equation when the ball hits the floor (exception).
A mathematically concise way to describe solution trajectories of
such systems in a well-defined matter would be to introduce
partial order between candidate sets of equations that the system
should satisfy and to take a maximally consistent element of the
partially ordered set (poset) of sets of constraints.  HydLa's design 
principle is exactly based on this idea.

\begin{figure}[t]
\hrule
\vspace{-6pt}
\begin{align*}
\texttt{INIT}  &\definedas \texttt{ht}\equals\texttt{10}
                \land \texttt{ht'}\equals\texttt{0}.\\[0pt]
\texttt{PARAMS}&\definedas \always(\texttt{g}\equals\texttt{9.8}
                \land \texttt{c}\equals{0.5}).\\[0pt]
\texttt{FALL}  &\definedas \always(\texttt{ht''}\equals\texttt{-g}).\\[0pt]
\texttt{BOUNCE}&\definedas \always(\texttt{ht-}\equals\texttt{0}
                \Rightarrow \texttt{ht'}\equals 
                \texttt{-c}\mathop{\texttt{*}}(\texttt{ht'-})).\\[6pt]
\texttt{INIT},\ &\texttt{PARAMS}, 
   (\texttt{FALL} \mathrel{\texttt{<<}} \texttt{BOUNCE}).
\end{align*}
\vspace{-12pt}
\hrule
\caption{A bouncing ball}
\label{fig:bouncingball}
\end{figure}
Figure~\ref{fig:bouncingball} shows the description of a bouncing
ball in HydLa.
The first four lines are the definition of constraint modules.
Constraint modules are program units which are combined to form
a set of constraints and to which priorities may be given.
In the right-hand side constraints, the postfix
`$\texttt{'}$' stands for a time derivative, the postfix minus sign
stands for the left-side limit of a trajectory, and $\always$ stands for
an \textit{always} temporal operator.
All the constraints stand for constraints at time 0.  However, since
the constraints other than $\texttt{INIT}$ start with $\always$,
they hold at all time points on and after time 0.
A constraint with an implication (such as $\texttt{BOUNCE}$)
is called a \textit{conditional constraint}.  A conditional constraint
prefixed by an $\always$ imposes its consequent exactly when
its antecedent (guard) holds.

The final line combines the four constraint modules.  A comma stands
for composition without priorities, while $\texttt{<<}$ gives a higher
priority to $\texttt{BOUNCE}$ than to $\texttt{FALL}$.  In this
example, all the four constraints are taken when the ball is in the
air, while
$\{\texttt{INIT},\texttt{PARAMS},\texttt{BOUNCE}\}$ will be taken
as the maximally consistent set when
the ball hits the floor because 
$\texttt{FALL}$ and $\texttt{BOUNCE}$ become inconsistent.

This example is known to exhibit a Zeno behavior,
an infinite number of discrete changes within a finite amount of time,
beyond which the simulation normally does not proceed.



\section{Basic HydLa}\label{sect:basichydla}

We consider the semantics of \emph{Basic HydLa} whose
syntax is shown in Fig.~\ref{fig:syntax}.
Basic HydLa simplifies HydLa~\cite{Ueda} as follows:

\begin{figure}[t]
\addtolength{\baselineskip}{3pt}
\begin{center}
\begin{tabular}{@{\quad}r@{~}r@{~}l@{}}
\hline\\[-3pt]
\textrm{(program)}& $P$ &$::=    (\textit{MS},\textit{DS}\,)$\\[2pt]
\textrm{(module set)}& \textit{MS} &$::= 
       \textrm{{poset of sets of }$M$}$\\[2pt]
\textrm{(definitions)}& \textit{DS} &$::=
       \textrm{{set of }$D$ {whose elements have
           different left-hand sides}}$\\[2pt]
\textrm{(definition)}& $D$ &$::= M \Leftrightarrow C$\\[2pt]
\textrm{(constraint)}& $C$ &$::= A\ \mid\ C \land C$
  $\ \ \mid\ G\Rightarrow C\ \mid\ \always C
       \ \mid\ \exists x.C$\\[2pt]
\textrm{(guard)}&      $G$ &$::=      A\ \mid\ G \land G$\\
$\begin{array}{r}\textrm{(atomic }\\[-2pt]
\textrm{constraint)}\end{array}\kern-5pt$&
                       $A$ &$::= E \mathop{\textit{relop}} E$\\
\textrm{(expression)}& $E$ &$::= \textrm{ordinary expressions}
     \ \mid\ E'\ \mid\ E-$\\[6pt]
\hline
\end{tabular}
\end{center}
\vspace{-6pt}
\caption{Syntax of Basic HydLa}
\label{fig:syntax}
\end{figure}

\begin{enumerate}
\item For each time point, HydLa chooses a consistent set of
  constraint modules that satisfies the priority constraint and that
  is maximal with respect to the set inclusion relation between
  constraint modules.
  More specifically, from a relative priority relation between
  constraint modules, HydLa first derives a poset
  whose elements are admissible (with regard to constraint priorities)
  sets of all the subsets of
  constraint modules \cite{Hirose},
  and then chooses a maximal consistent element.
  Basic HydLa does not handle this derivation but assumes that
  the ``(irreflexive) poset of sets of
    constraint modules'' is directly given in a program
    together with the definitions of constraint modules.
  \emph{Default} constraints such as the continuity of trajectories
  (frame axioms, see Section~\ref{sect:bibun})
  are to be explicitly specified within this poset.
  The constraints at the top of a constraint
  hierarchy should often be treated as \emph{required} constraints
  that \textit{must} be adopted, and
  whether to do so can be expressed explicitly within the poset.
\item Basic HydLa does not support the time shift (i.e. delay) operator
  `$\Hat{\;\;}$'.
  We can use the feature explained
  in the next item instead.
\item To enable dynamic creation of trajectories, Basic HydLa introduces
  an existential quantifier $\exists$ for local variable creation.
  This enables us to dynamically create a timer with which to represent a
  delay between the detection of some condition and the issue of a
  new constraint.
\item Basic HydLa does not support program definitions since they can
  be simply inlined.
\item For the same reason, Basic HydLa does not support the
  operator $\forall$ to generate a family of trajectories.
\end{enumerate}

We assume that a Basic HydLa program $(\textit{MS},\textit{DS}\,)$
satisfies $\bigcup\!\textit{MS} \subseteq \textrm{dom}(\textit{DS}\,)$,
where $\bigcup\!\textit{MS}$ is the set of modules appearing in
$\textit{MS}$ and $\textrm{dom}(\textit{DS}\,)$ is the set of
left-hand sides of $\textit{DS}$.
In the following, we consider a set $\textit{DS}$ of 
constraint module definitions as a function from module names to
constraints.

As shown in Fig.~\ref{fig:syntax}, we restrict the guard constraints
to atomic constraints and their conjunctions.
HydLa does not specify the class of constraints that can be described in
a program. In this sense, HydLa is a \emph{language scheme} that
parameterizes constraint systems.
The reason why we allow only $\always$ as a temporal operator is that
our syntax is targeted at the \textit{modeling} of systems.  Other temporal
operators such as $\eventually$ will be included in the specification
language when we construct a verification system that use HydLa
as a modeling language.


\section{Declarative Semantics of Basic HydLa}\label{sect:semantics}

As shown in Section~\ref{sect:overview}, the declarative semantics of
HydLa is defined as a relation meaning that a given trajectory (or
\textit{interpretation}) satisfies a program (or \textit{specification}).
The
information to be maintained by the declarative semantics depends on
design criteria such as what class of programs it deals with and what
degree of compositionality (i.e., the ability to compose the overall
semantics from the semantics of components) it aims at.  The semantics in
\cite{Ueda} dealt with programs containing no
$\always$ operators
in the consequents of conditional constraints.  Parameters and behaviors
of systems with a finite number of components and no delays can be
described by constraints with $\always$'s only in their prenex
positions.  When those programs contain 
conditional constraints, their consequents
hold exactly when the antecedents hold,
which means that a maximal
consistent set of constraints can be chosen at that time.

However, a constraint whose consequent includes an $\always$ leaves the
consequent as a candidate for choice even after the
corresponding antecedent ceases to hold.
If we have to judge 
which consequents of constraints should be chosen in the future
when the corresponding antecedents held, it would be a lookahead of the future.
Thus the choice of a maximal consistent set must be
performed not when constraints are discharged but
when the constraints are actually applied.  Therefore we further refine
our semantics in the following way.

First, we identify a conjunction of constraints with a set of constraints;
i.e., we view the syntax of a constraint in Fig.~\ref{fig:syntax}
as
$$C ::= \{A\}\,\mid\, C \cup C
  \,\mid\,\{G\Rightarrow C\}\,\mid\,\{\always C\}
     \,\mid\,\{\exists x.C\},
$$
and also allow an empty set.  By Skolemization, we recursively
eliminate existential quantifiers $\exists$ except for those occurring in
the consequents of conditional constraints.

Next, 
we consider constraint sets as functions of time.  For
example, a constraint $C$ that occurs in a program is regarded as a
function $C(0)=C$, $C(t)=\{\}$ $(t>0)$.

For a constraint $C(t)$ that is a function of time, the
$\always$-closure $C^*(t)$ is defined as a function that satisfies the
following properties:
\begin{itemize}
\item (Extension) $\forall t(C(t)\subseteq C^*(t))$;
\item ($\always$-closure)
$\forall t(\always a\!\in\!C^*(t) \Rightarrow
              \forall t'\ge t\,(a\subseteq C^*(t')))$;
\item (Minimality) For each $t$, $C^*(t)$ is the smallest set that
satisfies the above two conditions.
\end{itemize}
For $C=\{\texttt{f}\equals\texttt{0},
         \always\{\texttt{f'}\equals\texttt{1}\}\}$ for example,
we have
$C^*(0)=\{\texttt{f}\equals\texttt{0}, \texttt{f'}\equals\texttt{1},
          \always\{\texttt{f'}\equals\texttt{1}\}\}$,
$C^*(t)=\{\texttt{f'}\equals\texttt{1}\}$ $(t>0)$.

The constraint set that determines a solution trajectory of a HydLa
program may change over time for two reasons: one is that a
maximal consistent set may change; the other is that the consequent of a
conditional constraint is newly added when its antecedent holds.  The
choice of a maximal consistent set in the former case is performed
independently at each time point.  By contrast, when the
program has a constraint whose consequent begins with $\always$, 
whether the constraint is active or not
depends on whether its
antecedent has been activated \textit{in the past};
hence the state of a system should maintain the activation history
of the antecedents.
Therefore it is appropriate to consider a satisfaction relation stating that
a program
$P=(\textit{MS},\textit{DS}\,)$ is satisfied by a pair $\langle
\overline{x},Q\rangle$ of a solution trajectory
$\overline{x}=\overline{x}(t)$ and the constraint module definition
$Q=Q(M)(t)$ $(M\in\textrm{dom}(\textit{DS}\,))$ recording the
activation of antecedents.  We define this relation as shown in
Fig.~\ref{fig:semantics}.

\begin{figure}[t]
\def\qqquad{\qquad\quad}
\hrule
\vspace{-6pt}
\begin{align*}
&\kern-18pt
\rlap{$\langle \overline{x},Q\rangle\kern-1pt\models\kern-1pt
 (\textit{MS},\textit{DS}\,)
 \kern-2pt\definedas\kern-2pt
 \textrm{(i)}\kern-1pt\land\kern-1pt
 \textrm{(ii)}\kern-1pt\land\kern-1pt
 \textrm{(iii)}\kern-1pt\land\kern-1pt
 \textrm{(iv), where}$}\\[3pt]
\textrm{(i) }&\forall M\,(Q(M)=Q(M)^*);\\
\textrm{(ii) }&\forall M\,(\textit{DS}(M)^*\subseteq Q(M));\\
\textrm{(iii) }&
\forall t\,\exists E\!\in\!\textit{MS}\,(&       \textrm{(s0)}\\[-0pt]
  &\null\qquad (\overline{x}(t)\Rightarrow
               \{Q(M)(t)\mid M\in E\})&          \textrm{(s1)}\\[-0pt]
  &\land\ \lnot\,\exists \overline{x}'\,
                 \exists E'\!\in\!\textit{MS}\,(&\textrm{(s2)}\\[-0pt]
  &\null\qqquad \forall t'<t\,(\overline{x}'(t') =
                               \overline{x}(t'))&\textrm{(s2)}\\[-0pt]
  &\null\qqquad \land\ E \prec E'&               \textrm{(s2)}\\[-0pt]
  &\null\qqquad \land\ \overline{x}'(t) \Rightarrow
               \{Q(M)(t)\mid M\in E'\})&         \textrm{(s2)}\\[-0pt]
  &\land\forall d\,\forall e\,\forall M\!\in\!E\,(&\textrm{(s3)}\\[-0pt]
  &\null\qquad (\overline{x}(t)\Rightarrow d)\land
        ((d\Rightarrow e)\in Q(M)(t))
        \Rightarrow e\subseteq Q(M)(t)));&        \textrm{(s3)}\\
\textrm{(iv) }&\textrm{For each $M$ and $t$, $Q(M)(t)$ is the smallest set}\\[-0pt]
              &\textrm{that satisfies (i)--(iii).}
\end{align*}
\vspace{-12pt}
\hrule
\caption{Definition of $\langle \overline{x},Q\rangle\models P$}
\label{fig:semantics}
\end{figure}

The principle of the declarative semantics in Fig.~\ref{fig:semantics}
is the consistency-based adoption of constraints.  It requires that,
at each time point, a consistent set of constraint modules with
a maximal preference must be adopted and satisfied.

Condition (i) requires $Q(M)$ to satisfy the $\always$-closure property, and
Condition (ii) requires $Q(M)=Q(M)^*$ to be an extension of $\textit{DS}(M)^*$.
Now we look into Condition (iii) closely.
The order of the
quantifiers at Line (s0) allows $\overline{x}$ to choose, at each
time point, a different set of candidate modules from the constraint
hierarchy.
Line (s1) means that, at time $t$, $\overline{x}$ satisfies some set of
candidate modules in the constraint hierarchy.
Lines (s2) mean that there is no trajectory $\overline{x}'$ that
behaves exactly as $\overline{x}$ before $t$ and
satisfies a better candidate module set than $\overline{x}$ at $t$.
Lines (s3) mean that, when the antecedent of a chosen conditional
constraint holds, $Q$ is extended by expanding its consequent into the
definition of the corresponding module $M$ in $Q$.  If a member of
the consequent (regarded as a set of constraints) begins with
$\always$, it is further expanded by the $\always$-closure condition
(i).  Also, if it begins with $\exists$, the quantifier is eliminated
by using a Skolem function.  Condition (iv) requires the minimality.


\section{Examples}\label{sect:examples}

Using simple examples, we explain how the declarative semantics
actually defines solution trajectories and the constraint sets used
to determine them.

\medskip\noindent
\textbf{Example 1:} The first example shows how the arrival
of a monotonically 
increasing function value at a certain threshold is reflected
to another function with a delay.
\begin{align*}
P_1 &=(\textit{MS}_1,\textit{DS}_1\,)\\[-0pt]
\textit{MS}_1 &= (\{\{\texttt{A},\texttt{C}\},
                    \{\texttt{A},\texttt{B},\texttt{C}\}\},
                  \{\{\texttt{A},\texttt{C}\}
               \prec\{\texttt{A},\texttt{B},\texttt{C}\}\})\\[-0pt]
\textit{DS}_1 &= \{\,
\texttt{A} \definedas \texttt{f}\equals\texttt{0}\land 
                      \always(\texttt{f'}\equals\texttt{1}),\\[-0pt]
 &\qquad \texttt{B} \definedas \always(\texttt{g}\equals\texttt{0}),\\[-0pt]
 &\qquad \texttt{C} \definedas 
\always (\texttt{f}\equals\texttt{5} \Rightarrow \exists \texttt{a}\,.\,
(\texttt{a}\equals\texttt{0} \land \always(\texttt{a'}\equals\texttt{1})
     \land \always (\texttt{a}\equals\texttt{2}
                    \Rightarrow \texttt{g}\equals\texttt{1})))\}
\end{align*}
Here,
$\texttt{f}$ is a function that expresses the elapsed time,
$\texttt{a}$ is a timer invoked by $\texttt{f}\equals\texttt{5}$ as the
trigger, and $\texttt{g}$ is a pulse function that is usually 0 but
momentarily becomes 1 two seconds after the invocation of the timer.  The
solution trajectory $\overline{x}$ expresses those behaviors of
$\texttt{f}$, $\texttt{a}$ (whose Skolem function is also called
$\texttt{a}$ here), and $\texttt{g}$.  Now we see all the constraints
$Q(*)(t)$ $=$ $\bigcup\{Q(M)(t)\mid M\in \textrm{dom}(DS_1)\}$ that
are stored in $Q$.  At $0<t<5$, $Q(*)(t)$ consists of
$\texttt{f'}\equals\texttt{1}$, $\texttt{g}\equals\texttt{0}$, 
and the constraint $\texttt{C}$ with
the leftmost $\always$ removed. At $t=5$, $\texttt{a}\equals\texttt{0}$,
$\always(\texttt{a'}\equals\texttt{1})$, $\texttt{a'}\equals\texttt{1}$, 
$\always
(\texttt{a}\equals\texttt{2}\Rightarrow \texttt{g}\equals\texttt{1})$, 
and $\texttt{a}\equals\texttt{2}\Rightarrow
\texttt{g}\equals\texttt{1}$ are added to them.  
At $5<t<7$, $\texttt{a}\equals\texttt{0}$,
$\always(\texttt{a'=1})$, and $\always (\texttt{a=2}\Rightarrow
\texttt{g}\equals\texttt{1})$ are removed.  
At $t=7$, $\texttt{g}\equals\texttt{1}$ replaces
$\texttt{g}\equals\texttt{0}$.  
At $t>7$, $\texttt{g}\equals\texttt{1}$ is replaced by
$\texttt{g}\equals\texttt{0}$ again; the other constraints that remain are
$\texttt{f'}\equals\texttt{1}$, $\texttt{a'}\equals\texttt{1}$ 
and the two conditional constraints.

\medskip\noindent
\textbf{Example 2:} The declarative semantics presented in the
previous section disallows the propagation of constraints to the past.
This may be obvious from the
construction of the semantics, but we confirm it by using an example
since it is an important property.
\begin{align*}
P_2 &= (({\cal P}(\{\texttt{D},\texttt{E},\texttt{F}\}),\subsetneq),
        \textit{DS}_2\,)\\[-0pt]
\textit{DS}_2 &= \{\,
\texttt{D} \definedas \texttt{y}\equals\texttt{0},\\[-0pt]
&\null\qquad \texttt{E} \definedas
              \always (\texttt{y'}\equals\texttt{1} 
                       \land \texttt{x'}\equals\texttt{0}),\\[-0pt]
&\null\qquad \texttt{F} \definedas
              \always (\texttt{y}\equals\texttt{5}
                       \Rightarrow \texttt{x}\equals\texttt{1})\}
\end{align*}
$P_2$ leaves the initial value of $\texttt{x}$ undefined.  We check
whether the constraint $\texttt{x=1}$ imposed by $\texttt{F}$ at
$t=5$ propagates to the past by the effect of $\texttt{x'}\equals\texttt{0}$ in
$\texttt{E}$.  We consider the following three cases as candidates for
solution trajectories.

\begin{enumerate}
\item $\texttt{y}(t)=t$ $(t\ge 0)$ and $\texttt{x}(t)=1$ $(t\ge 0)$
satisfy all the constraints $\texttt{D}$, $\texttt{E}$, and
$\texttt{F}$ at all times.
\item $\texttt{y}(t)=t$ $(t\ge 0)$ and $\texttt{x}(t)=2$ $(t\ge 0)$
satisfy all the constraints except at $t=5$ and satisfy
$\texttt{D}$ and $\texttt{E}$ at $t=5$.
\item $\texttt{y}(t)=t$ $(t\ge 0)$, $\texttt{x}(t)=2$ $(t<5)$, and
$\texttt{x}(t)=1$ $(t\ge 5)$ satisfy all the constraints except at
$t=5$ and satisfy $\texttt{D}$ and $\texttt{F}$ at $t=5$.
\end{enumerate}

Case~1 is a solution since it obviously satisfies the maximality.
Cases~2 and~3
obviously satisfy the maximality except at $t=5$, and there are no
better solutions than these.  Neither of them is worse than the other
at $t=5$, and there are no other solutions that satisfy all the
constraints; hence both of them are maximal.
In other
words, any of Cases~1 to~3 is a result of ``extending a solution
along the time axis so the maximality is satisfied,'' and is therefore a
solution to $P_2$.


\section{Discussions on the Specification and the Semantics of the
  Language}\label{sect:discussions}

\subsection{Differential Constraints}\label{sect:bibun}

The basic principle of HydLa to utilize existing
mathematical and logical notations as much as possible suggests that
the precise meaning of the notations should also conform to mathematical
conventions.
For example, at the time point where a piecewise continuous trajectory
causes a discrete change, we do not consider the trajectory
differentiable even if 
it is differentiable both from the left and the right,
and we do not deactivate the differential constraints at
that time point.
We also assume only the right continuity and right differentiability
at the initial time.

For the reasons above, 
the priority of the differential constraints 
of a piecewise continuous function
should in general be lower than
that of the constraints describing discrete changes.
On the other hand, for a continuous trajectory after a
discrete change to be well-defined as an initial value problem
of an ordinary differential equation,
we need to assume the right continuity at the time of
the discrete change.
Since the differential constraints are deactivated when a discrete
change occurs, we also require left continuity to be able to decide
the value of a trajectory. 
Accordingly, HydLa assumes both the right and the left
continuity of trajectories described by differential constraints.
Since these two continuity constraints are automatically entailed whenever
a trajectory is differentiable, we assume them separately with
a priority higher than differential constraints.

\subsection{Expressive Power of HydLa}

Although the primary purpose of HydLa is to describe piecewise
continuous trajectories,
we can define various trajectories or sets of trajectories using
HydLa's constraints and constraint hierarchies.

\subsubsection{Trajectories defined without using differential equations}

HydLa allows us to describe trajectories without
using differential constraints.
For example, 
a drifting parameter can be described by 
a constraint $\always(\texttt{0.9}\mathop{\texttt{<}}\texttt{a} 
                  \land \texttt{a}\mathop{\texttt{<}}\texttt{1.1})$, which
represents the set of all trajectories whose range is $(0.9,1.1)$.

Note that a trajectory defined by the above constraint may not be
continuous. Hence, a trajectory defined by
$\texttt{f}\equals\texttt{0}\land \always(\texttt{f'}\equals\texttt{1})$
is not guaranteed to satisfy $\texttt{f}\equals\texttt{a}$
between time $0.9$ and $1.1$.
By adding a constraint $\always(\texttt{a'}\equals\texttt{b})$
(we do not add any
constraint on $\texttt{b}$), $\texttt{a}$ stands for a set of all
continuous and differentiable trajectories whose range is $(0.9,1.1)$,
and is guaranteed to intersect with $\texttt{f}$.

A pulse function is another example defined without
differential constraints. 
An example of a pulse function is \texttt{g} of Example~1 
(Section~\ref{sect:examples}).
Pulse functions play a significant role in representing the occurrences
of events.
Since pulse functions are not right-continuous at the time of
discrete changes, we conjecture that
a trajectory after the discrete change cannot be defined directly by a
differential equation.
The following example shows that one of the attempts to 
define a pulse function \texttt{b} fails:
\begin{align*}
P_3 &= (\textit{MS}_3,\textit{DS}_3\,)\\[-0pt]
\textit{MS}_3 &= (\{\{\texttt{G},\texttt{J}\},
                    \{\texttt{G},\texttt{H},\texttt{J}\}\},
                  \{\{\texttt{G},\texttt{J}\}
               \prec\{\texttt{G},\texttt{H},\texttt{J}\}\})\\[-0pt]
\textit{DS}_3 &= \{\,
        \texttt{G} \definedas
                   \texttt{a}\equals\texttt{0} 
                   \land \texttt{b}\equals\texttt{0}
                   \land \always(\texttt{a'}\equals\texttt{1}),\\[-0pt]
&\qquad \texttt{H} \definedas \always(\texttt{b'}\equals\texttt{0}),\\[-0pt]
&\qquad \texttt{J} \definedas 
                   \always(\texttt{a-}\equals\texttt{1} 
                     \Rightarrow \texttt{b}\equals\texttt{1})
             \land \always(\texttt{b-}\equals\texttt{1} 
                     \Rightarrow \texttt{b}\equals\texttt{0})\}
\end{align*}
Based on the discussion in Section~\ref{sect:bibun}, between two sets of
constraint modules in $\textit{MS}_3$, there exist several sets
with additional constraints on the continuity including
the right continuity of \texttt{b}.
At $t=1$, the set $\{\texttt{G},\texttt{H},\texttt{J}\}$ is not satisfiable but
$\{\texttt{G},\texttt{J}\}$ with the right continuity of \texttt{b}
is satisfiable, and $\texttt{b}(1)=1$ holds from the first
constraint of \texttt{J}.
However, then, the greatest lower bound of the time when
the guard of the second constraint of \texttt{J} holds
is $t=1$.
The consequent of the constraint $\texttt{b}(t)=0$ is thus
activated at $t>1$ and
contradicts the right continuity.
Now suppose we drop the assumption of the right continuity.
Then it turns that $\texttt{b}(t)=c$ ($t>1$) is consistent for all
$c\ne 1$.
Therefore, although there exists a solution trajectory, HydLa fails to
guarantee its uniqueness.

\subsubsection{Zeno behaviors}

Let us reinvestigate the bouncing ball example in
Section~\ref{sect:overview} based on the declarative semantics of
Section~\ref{sect:semantics}.
Although the program in Fig.~\ref{fig:bouncingball} specifies a
unique solution trajectory until the Zeno time,
after that, it allows a trajectory that falls through the floor.
We need some additional rules to specify the behavior after the Zeno
time~\cite{Zeno}.
In HydLa, we can define it as 
$\always(\texttt{ht-}\equals\texttt{0}\land
\texttt{ht'-}\equals\texttt{0}\Rightarrow
\always(\texttt{ht}\equals\texttt{0}))$,
though checking the guard condition would
need a special simulation method, e.g., in~\cite{OhnoZeno}. 

The following program shows another method for detecting the Zeno time.
It checks the convergence of a function $\texttt{vmax}$ that holds
the velocity at the last bounce. 
\begin{align*}
&\kern-30pt\always(\texttt{vmax'}\equals\texttt{0})
                   \mathrel{\texttt{<<}}\\[-0pt]
&\always(\texttt{ht'-}\mathop{\texttt{!=}}\texttt{ht'}
                   \Rightarrow \texttt{vmax}\equals\texttt{ht'})\\[-0pt]
\land\ &\always(\texttt{vmax-}\equals\texttt{0}
                   \Rightarrow \always(\texttt{ht}\equals\texttt{0}))
\end{align*}
This example shows that the left limit operator `$\texttt{-}$' is
also useful for a function that only causes discrete changes.


\section{Conclusions and Future Work}

This paper presented the declarative semantics of HydLa, a hybrid
constraint language with hierarchical structure, described its
mechanisms and consequences by means of examples, and discussed the
language features and expressive power.

The semantics given in this paper regards trajectories as functions
over time.  On the other hand, the theory of hybrid systems often
adopts hybrid time that allows more than one discrete change at a
single time point \cite{hybridtime}.  One of the motivations of hybrid
time is to model computation involving multiple steps at the time of a
single discrete change.  However, because HydLa is constraint-based,
such evolution can be represented as constraint propagation rather
than state changes.  Another motivation of hybrid time is to deal with
the stability and convergence of trajectories in a declarative
framework.  This would require the extension of our
semantics with hybrid time, which is a topic of future work.

We are currently working on the formulation and its implementation of
a simulation algorithm corresponding to our declarative semantics.
The resulting system is planned to exploit the flexibility of
constraint programming and its affinity to interval computation.

\medskip\noindent
{\bf Acknowledgments}\
The design of HydLa has received constant feedback from the
theoretical and implementation work and daily discussions of the past
and present members of the HydLa project.  This research is partially
supported by Grant-In-Aid for Scientific Research
((B) 20300013), JSPS, Japan.


\end{document}